\documentclass[preprint,12pt]{elsarticle}

\usepackage{graphicx}
\usepackage{dcolumn}
\usepackage{bm}
\usepackage{amsfonts}
\usepackage{amsmath}
\usepackage{amssymb}
\usepackage{latexsym}
\usepackage{textcomp}
\usepackage{tikz}

\PassOptionsToPackage{normalem}{ulem}
\usepackage{ulem}
\providecolor{added}{rgb}{0,0,1}
\providecolor{deleted}{rgb}{1,0,0}

\journal{Applied Surface Science}

\begin{document}
\def\myfrac#1#2{\frac{\displaystyle #1}{\displaystyle #2}}

\begin{frontmatter}

\title{On femtosecond laser shock peening of stainless steel AISI 316}

\author{Jan S. Hoppius}
\ead{hoppius@lat.rub.de}
\address{Chair of Applied Laser Technologies, Ruhr-Universit\"at Bochum, Universit\"atsstra\ss e~150, 44801 Bochum, Germany}

\author{Lalit M. Kukreja}
\address{Epi-Knowledge Foundation, C 2-4/1:3, Sector-4, Vashi, Navi Mumbai - 400 703, India}

\author{Marina Knyazeva}
\address{Chair of Materials Test Engineering, TU Dortmund University, Baroper Str. 303, D-44227 Dortmund, Germany}

\author{Fabian P\"ohl}
\address{Chair of Materials Testing, Ruhr-Universit\"at Bochum,	Universit\"atsstra\ss e~150, 44801 Bochum, Germany}

\author{Frank Walther}
\address{Chair of Materials Test Engineering, TU Dortmund University, Baroper Str. 303, D-44227 Dortmund, Germany}

\author{Andreas Ostendorf}
\address{Chair of Applied Laser Technologies, Ruhr-Universit\"at Bochum, Universit\"atsstra\ss e~150, 44801 Bochum, Germany}
 
\author{Evgeny L. Gurevich}
\ead{gurevich@lat.rub.de}
\address{Chair of Applied Laser Technologies, Ruhr-Universit\"at Bochum, Universit\"atsstra\ss e~150, 44801 Bochum, Germany}

\date{\today}

\begin{abstract}

In this paper we report on the competition in metal surface hardening between the femtosecond shock peening on the one hand, and formation of laser-induced periodic surface structures (LIPSS) and surface oxidation on the other hand. Peening of the stainless steel AISI 316 due to shock loading induced by femtosecond laser ablation was successfully demonstrated. However, for some range of processing parameters, surface erosion due to LIPSS and oxidation seem to dominate over the peening effect. Strategies to increase the peening efficiency are discussed.

\end{abstract}

\begin{keyword}
femtosecond laser \sep shock peening \sep LIPSS \sep surface oxidation
\end{keyword}

\end{frontmatter}


\section{Introduction}

The stainless steel AISI 316 is a high corrosion resistance steel, which finds applications in medicine, turbines and aerospace components. It is imperative to enhance the lifetime of the components operating under extreme conditions of dynamic stress, temperature and corrosive nature of the ambiance. One way to accomplish this is by peening. A superior and contemporary method of peening is by shock loading due to pulsed laser ablation \cite{ding2006laser,Liao2016}. In order to increase the effect, the laser light is usually focused onto a sacrificial layer, which is attached to the surface and surrounded by a confining medium, usually water. Conventionally, millisecond and nanosecond laser pulses have been applied for this purpose. However, the ultrafast (or femtosecond) lasers provide very different conditions for the shock peening, which can be advantageous for some peening applications. Indeed, the pressure of the shock waves of up to 100-1,000 GPa can be achieved with the pulse energy in the mJ range in the case of femtosecond laser pulses, whereas the nanosecond lasers can deliver only 1-10 GPa pressure with the pulse energy in the Joule range \cite{age2016}. The depth of the peening grows with the pulse energy, hence femtosecond shock peening should improve hardness more efficiently than traditional nanosecond peening if only a thin surface layer is affected, e.g., for thin layers and micro parts. 

Several successful experimental studies on femtosecond laser peening of metals have recently been carried out \cite{nak2009, nak2010, Lee2011, Ye014, dut2016}. These studies demonstrated an increase in the sample hardness due to femtosecond laser shock peening of metals, although usually it doesn't exceed an increase of 20\%. The reason of such a moderate (in comparison to the 100-fold stronger amplitude of the shock wave) increase has not been discussed in the literature to the best of our knowledge. In the present studies it was found that the shock wave is not the only process, which is initiated on the sample surface by the laser. The generation of LIPSS due to direct interaction of the metal surface with the femtosecond laser light (1), and surface oxidation due to interaction of the laser produced plasma with the metal surface (2) are side effects interfering with the surface hardening.
Direct interaction of laser light with metals, dielectrics and semiconductors is known to generate laser-induced periodic surface structures (LIPSS) or ripples \cite{bir1965,BonseRev}. These ripples manifest themselves as a periodically modulated profile of the sample surface. They interfere with the local mechanical properties of the surfaces and the hardness measurements, because the period of the LIPSS is usually in the sub-micrometer range. In this paper, the results of investigations for the search of the optimal parameters for the shock peening in order to prevent the LIPSS formation and accomplish the pure shock peening effect are presented.
Transparent dielectrics are usually used as confining medium. Unlike nanosecond laser, femtosecond laser pulses can interact with such materials due to extremely high peak fluence. Self-focusing can increase the intensity in the confining media above the breakdown threshold due to filamentation of the laser beam \cite{Yariv1978}. It is possible that the confining medium will be ionized by the laser radiation due to multiphoton absorption or tunnel ionisation and the plasma in the environment will be ignited \cite{Schaffer2001,OptBreakdown,Bulgakova2010}. Interaction between the sample surface and the environmental plasma will lead to surface oxidation or carbidization \cite{Kanitz2017}.

\section{Experimental Setup}

\begin{figure}[h!]
 \centerline{\includegraphics[width=9cm]{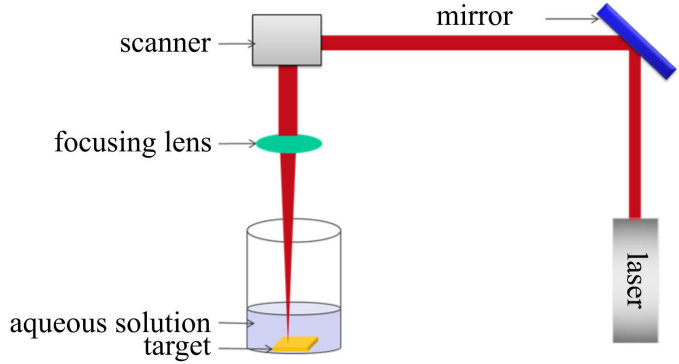}}
\caption{Schematic representation of the experimental setup.}
\label{setup}
\end{figure}

The experimental setup used in this study is schematically shown in Fig.~\ref{setup}. The sample was immersed in a container with deionized water which served as confining medium. In some experiments the sample surface was covered by a 0.15 mm thick sticky tape, which was used as sacrificial layer. 

Laser IR-radiation ({\it Spitfire Ace}, produced by {\it Spectra Physics}, central wavelength $\lambda$=800 nm, pulse duration less than 100 fs, output energy in the range of $E_p$=0.01-0.75 mJ) was focused by an F-Theta objective resulting in a spot diameter of 10 \textmu m in the focal plane. The repetition rate was fixed at 5 kHz. The position of the focal spot on the sample surface was controlled by a galvanometer scanner ({\it SCANcube10}, produced by {\it SCANLAB}).

Plasma breakdown as well as subsequent heating and vaporization lead to the formation of cavitation bubbles and long living micro bubbles. As light scattering on these bubbles decreased the experimental reproducibility, a water flow was used to rapidly remove them from the processing area.

As a sample we used AISI 316 stainless steel purchased from {\it Good Fellow}. In the first set of experiments the sample surface was not covered with any sacrificial layer, whereas in the second set we covered the surface with an approximately 150\,\textmu m thick sticky tape.

\section{Results and discussion}
\subsection{Experiments without additional sacrificial layer}
Application of a sacrificial layer seemed to be questionable in the case of femtosecond laser peening, because of the fact that the large pressure in the shock wave is achieved by a low pulse energy (milli- or sub-milli Joule range). Due to the low pulse energy, the penetration depth of the shock wave is comparatively small. For example, it was impossible to detect any influence of the shock wave by optical analysis of the sample cross-section, hence the affected length must be less than one micrometer. This estimation agrees with the results of the numerical simulations \cite{iva2003}, which predict the shock wave relaxation on the sub-micrometer scale. Hence, experiments reported in this section were done without sacrificial layers.

\begin{figure}[h!]	
	\centerline{
		\includegraphics[width=13.4cm]{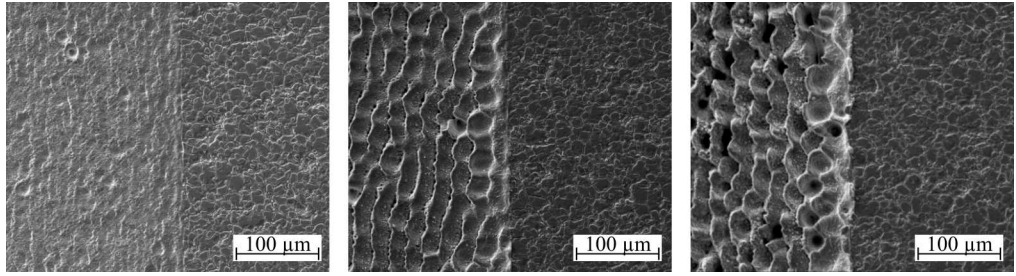}}
	\caption{SEM images of the laser-irradiated surfaces. The laser pulse energy was (left) $E_p$ = 200 \textmu J, (center) $E_p$ =400 \textmu J, (right) $E_p$ =600 \textmu J. The virgin surfaces are shown on the right-hand side.}
	\label{SEM}
\end{figure}

Figure~\ref{SEM} shows the SEM images of AISI 316 surfaces irradiated with 200, 400 and 600 \textmu J pulse energy (255, 510 and 765 J/$cm^2$ laser fluence in the focal plane) in water. The surface topography was obviously strongly dependent on this parameter. Clear traces of the laser ablation can be seen in all samples irradiated with more than 400 \textmu J. The observed craters with an average diameter of 22 \textmu m are larger than the focal spot size and their arrangement show deviations from the vertical processing lines. Waves at the water surface and micro bubbles on top of the metal sample scatter, deflect and defocus the laser beam and lead to this non-uniform surface profile.
The EDX analysis of these areas indicated 25 to 36 atomic percent of oxygen on the surface, which means oxidation by the laser-induced plasma in water. 

\begin{figure}[h!]\centerline{	
		\includegraphics[width=13.4cm]{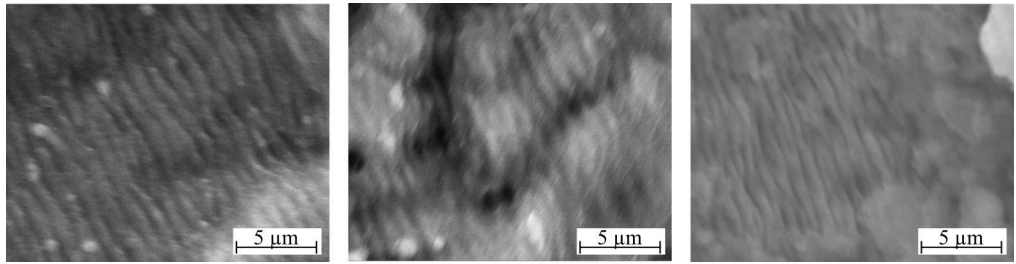}}
	\caption{High resolution SEM images of the laser-irradiated surfaces. The laser pulse energy was (left) $E_p$ = 200 \textmu J; (center) $E_p$ =400 \textmu J; (right) $E_p$ =600 \textmu J. LIPSS can be clearly observed.}
	\label{SEMnear}
\end{figure}  

A closer look at the laser irradiated surfaces, as shown in Fig.~\ref{SEMnear}, revealed the formation of laser induced periodic surface structures: a periodic set of thin metal lamelas, which thickness is several tens of nanometers and the period is less than one micrometer. The LIPSS in Fig.~\ref{SEMnear} indicate that in this regime the surface is strongly influenced directly by the laser light and the combined surface treatment takes place. 

From these studies it is clear that even though the residual compressive stress could be generated in the metal along the depth below the surface, as it is reported by Majumdar {\it et al.} \cite{dut2016}, the resulting surface itself posed complex problems. For the effective application of the methodology of femtosecond laser peening it is imperative to investigate these complex problems in depth to generate comprehensive insight on this subject. Two such problems that were identified are the LIPSS and surface oxidation due to the direct interaction of the femtosecond laser light and the interaction of the laser produced plasma with the metal surface, respectively. To the best of author's knowledge these surface phenomena have not yet been investigated or reported in the literature for their effects on the mechanical properties of the femtosecond laser peened AISI 316 surfaces.

\begin{figure}[h!]\centerline{\includegraphics[width=7cm]{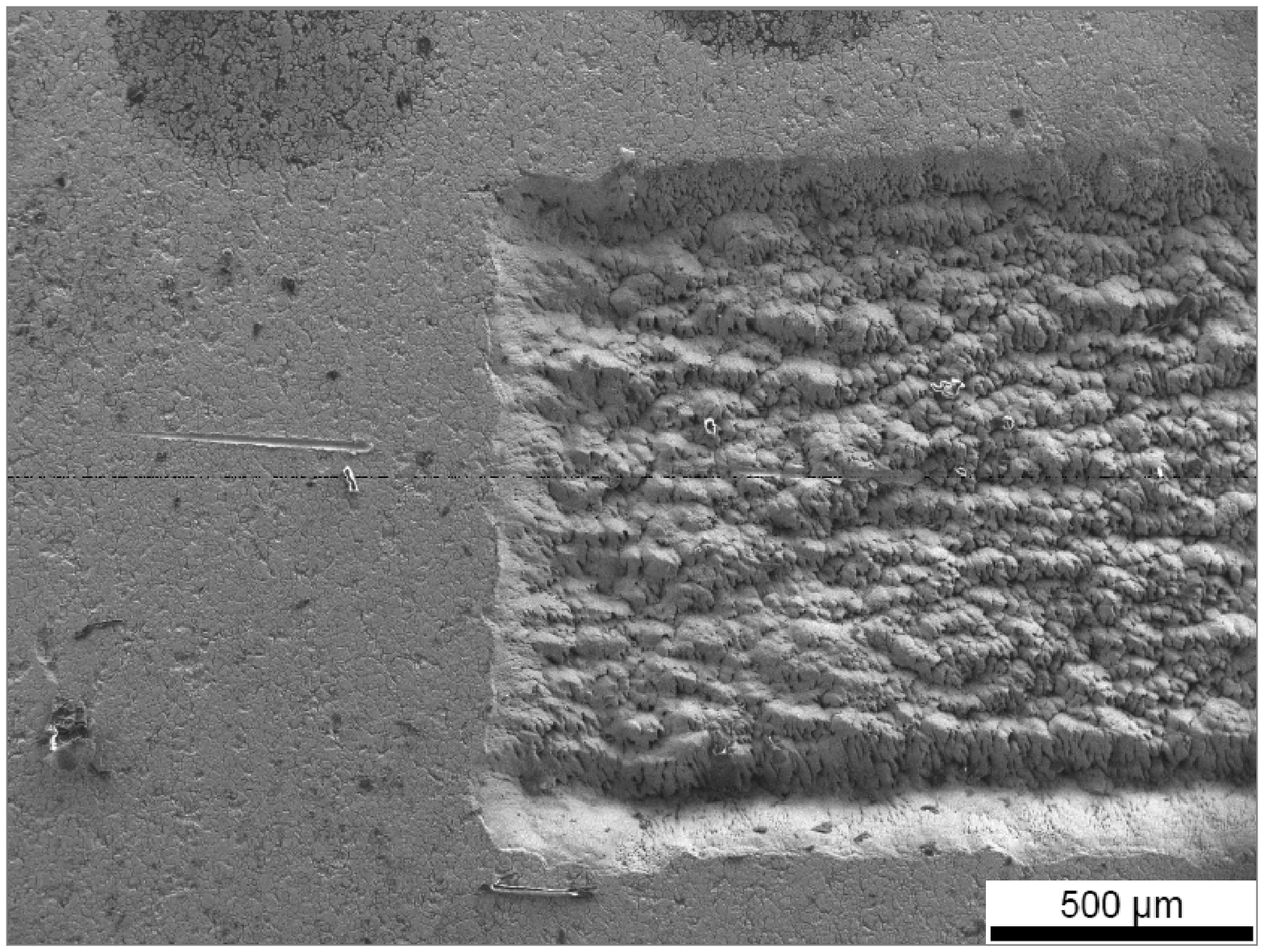}
		\vspace{0.5cm}
		\includegraphics[width=7cm]{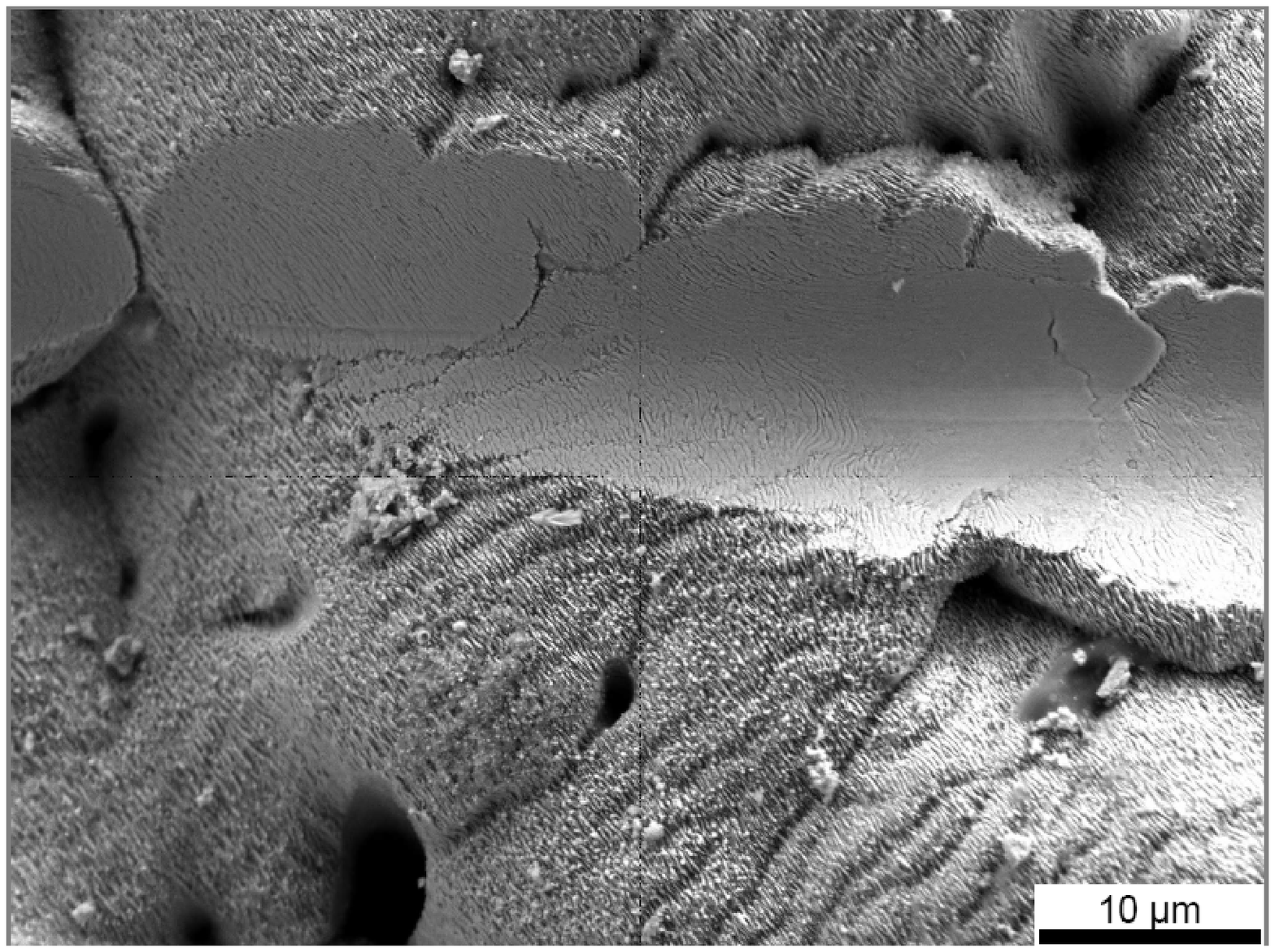}}
	\caption{Overview of indenter scratch in beside and in laser irradiated (800\,\textmu J pulse energy) area (left)  and detailed image of indenter scratch through LIPSS (right)}
	\label{scratchpic}
\end{figure}

\begin{figure}[h!]
	\centerline{\includegraphics[width=7cm]{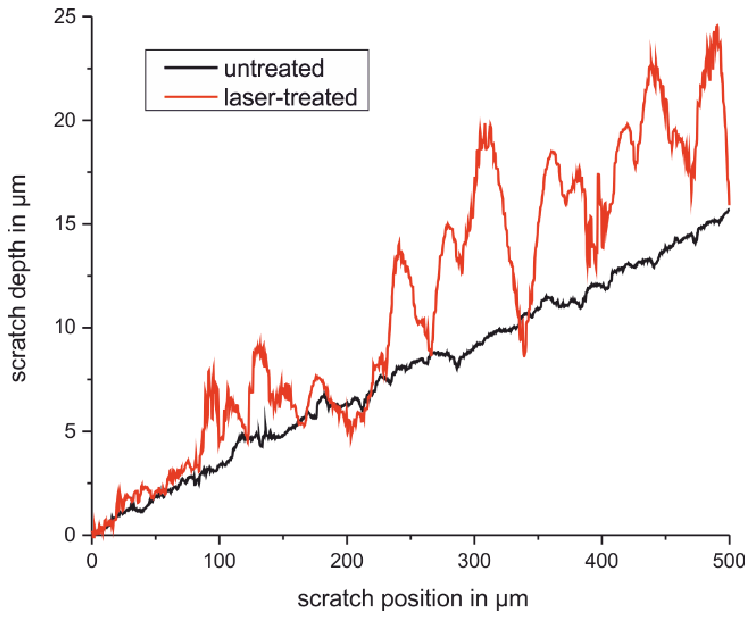}	
		\vspace{0.5cm}	
		\includegraphics[width=7cm]{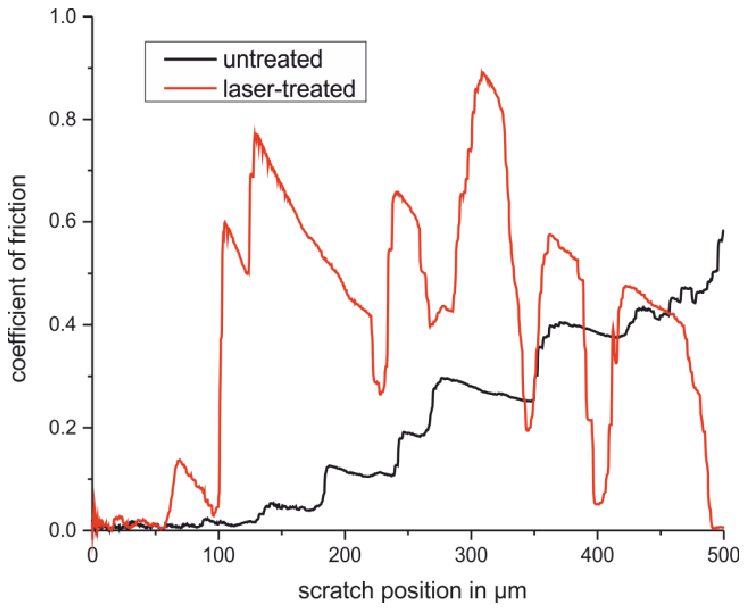}}
	\caption{Scratch depth (left) and coefficient of friction (right) in the laser irradiated and the virgin area of AISI 316 for linearly increasing force.}
	\label{scratchgraph}
\end{figure}  

\begin{figure}[h!]
	\centerline{\includegraphics[width=7cm]{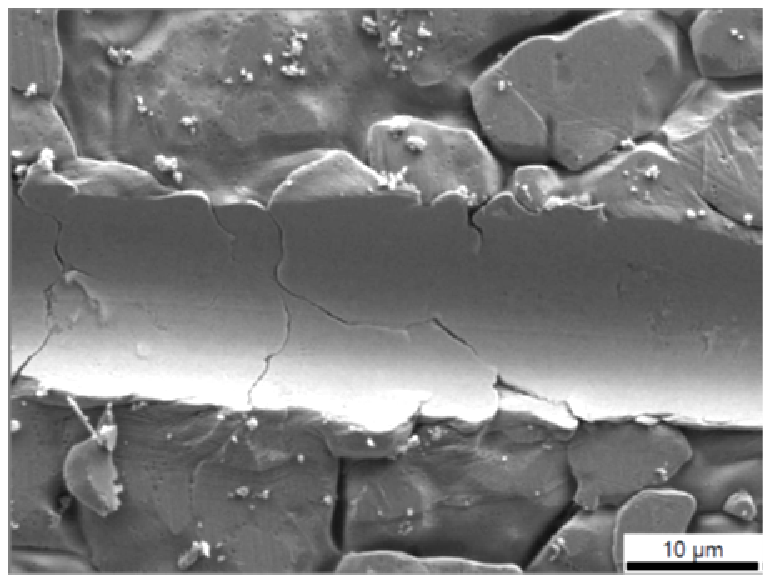}
	\vspace{0.5cm}	
	\includegraphics[width=7cm]{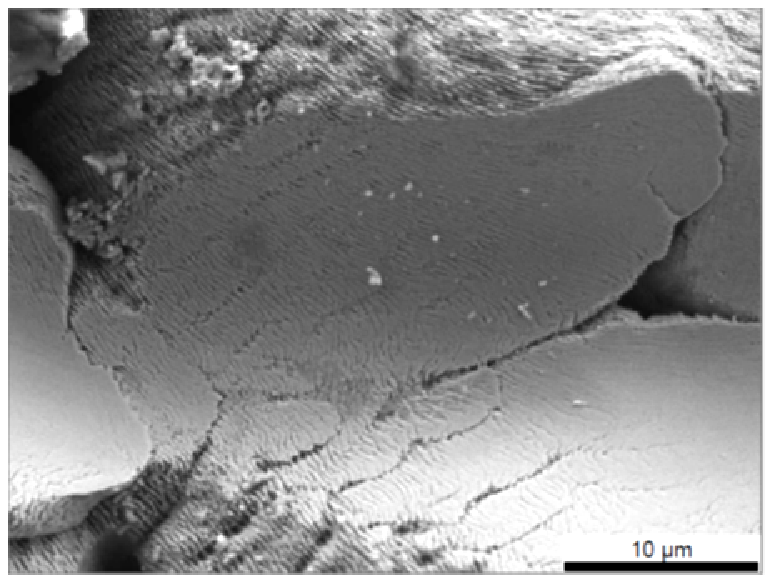}}
	\caption{Comparison of scratch traces of virgin (left) and of laser irradiated (right) areas.}
	\label{scratchtraces}
\end{figure}

The high degree of roughness due to ablation on the micrometer scale and LIPSS  formation on the sub-\textmu{}m scale (see Fig.~\ref{scratchpic} right) makes nano-indentation measurements impossible. To compare the mechanical properties of the laser-treated and untreated AISI 316 scratch characteristics and coefficient of friction were determined ({\it Nano-Scratch Tester}, produced by {\it CSM Instruments}, 10\,\textmu m sphero-conical diamond tip, constantly increasing force from 3\,mN to 503\,mN applied to the probe). In the laser-processed part of the sample (see Fig.~\ref{scratchpic} left) one can see an ablated area with redeposited nanoparticles. The scratch depth and coefficient of friction in both virgin and laser irradiated areas are presented in Fig.~\ref{scratchgraph}.  The average indentation depth of the scratches in laser irradiated areas is larger and significantly more fluctuating than in untreated areas. This may be attributed to the reduced strength caused due to morphology of ablation and LIPSS or surface oxidation observed in the laser irradiated area \cite{Kanitz2017} of AISI 316. The tip can easily be pressed into this periodic structure during the scratch test, as it can be seen in the Fig.~\ref{scratchtraces}~left. The coefficient of friction is deducted from the tangential force needed to scratch the surface. In the irradiated area it is higher because a larger amount of material has to be relocated by the scratch tip. Due to the increased roughness, the fluctuations in the friction coefficient are larger and the overall strength against scratches is lower. This unwelcome hardness-reducing effect dominates over the hardness increase due to peening here. Thus, since the overall effect on the surface hardness is negative in experiments without sacrificial layer, the latter will be used in the following experiments.

\subsection{Experiments with sacrificial layer}

\begin{figure}[h!]
 \centerline{\includegraphics[width=14cm]{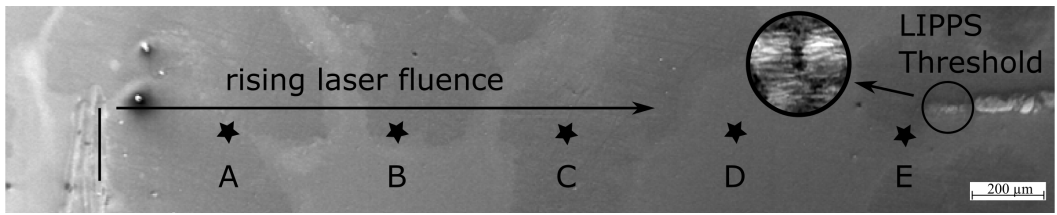}}
\caption{AISI 316 sample with adhered sacrificial tape was processed with varying laser focus position. The resulting laser fluence varied from A to E and corresponds 5\,J/$cm^2$, 6.8\,J/$cm^2$, 9.6\,J/$cm^2$, 14.7\,J/$cm^2$ and 25\,J/$cm^2$, respectively. At higher fluence the tape was completely ablated and LIPSS appeared.}
\label{gradientexp}
\end{figure}

Lee {\it et al.} \cite{Lee2011} used  approximately 27\,\textmu m zinc coating as sacrificial layer. From the practical point of view, coating of real metallic samples with other metals is hard to implement, so we sticked adhesive tape on top of the stainless steel. There are two reasons to use the sacrificial layer: (1) to protect the surface from LIPSS formation and from oxidation. (2) to increase the pulse energy above the ablation threshold and still keep the sample surface intact. The thickness of the tape is larger than both the ablation depth and the depth of the shock wave propagation, hence a single pulse on the covered sample did not affect the metal surface. 
The number of pulses per site (or pulse overlap) must be chosen in such a way, that the last pulse removes the sacrificial layer, or in other words, is absorbed in the sacrificial layer of a negligible thickness. This will prevent the direct interaction between laser light and the surface but keep the shock wave close enough to the sample surface to maximize the peening action. 
The second parameter controlling the peening efficiency is the pulse energy, which defines as well the depth ablated by each laser pulse, as the shock wave amplitude. 

Therefore, we processed 5$\times$0.5 mm$^2$ area with multiple spots per site, where laser pulse energy (800\,\textmu J), pulse repetition rate (5\,kHz) and lateral feed rate (1\,mm/s) were fixed while the axial laser focus position was shifted from 1\,mm below the metal surface to the top of the sticky tape. Accordingly, the interacting laser fluence on the sample surface increases in the direction of the arrow in Fig.~\ref{gradientexp}. The marked positions A to E correspond to 5\,J/$cm^2$, 6.8\,J/$cm^2$, 9.6\,J/$cm^2$, 14.7\,J/$cm^2$ and 25\,J/$cm^2$, respectively. Directing the axial laser focus further to the top, the tape was completely ablated and LIPSS appeared at the metal surface. Since the laser fluence that interacts with the tape is dependent on the already ablated tape and the progress of focal plane shifting, the ablation depth per pulse is not constant. An average ablation depth of 18\,\textmu m per shot was determined if the focus plane is on the metal surface.

\begin{figure}[h!]
 \centerline{\includegraphics[width=9cm]{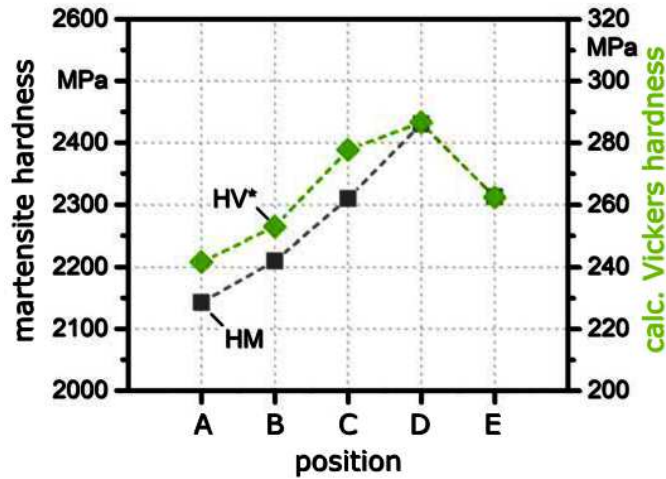}}
\caption{Martensite hardness (HM) and calculated Vickers hardness (HV) with increasing laser fluence. Points A-E correspond to the positions marked in Figure~\ref{gradientexp}, whereas the point E is close to the onset of LIPSS.}
\label{hardness}
\end{figure}

The microindentation hardness from position A to E, plotted in Fig.~\ref{hardness}, were performed with a fixed penetration depth of 800\,nm and reveal a functional correlation between focus position and hardness from position A to D. Before the total tape ablation is reached, the martensite hardness increases with the laser fluence. However, the hardness decreases as fluence is so high that the sacrificial layer is fully removed before the end of the processing. This can be assigned either to different mechanisms discussed in the previous section or to formation of under-surface defects. Indeed, the expected pressure in our experiments exceeds the Hugoniot elastic limit, which is approximately 1\,GPa for 316L stainless steel \cite{Peyre}. It is not clear whether sub-surface defects can be formed and deteriorate the hardness if the pressure is much higher than the elastic limit.

\section{Conclusions}

Femtosecond shock peening of AISI 316 is able to increase the surface hardness if two competing processes (surface oxidation and LIPSS formation) are inhibited. An easy way to achieve this is to use a sacrificial layer. However, application of the sacrificial layer restricts the processing parameters, especially the number of pulses per site and the pulse fluence. They should be chosen in agreement with the thickness of the layer in such a way, that the sacrificial material is fully ablated only by the last laser pulse. In this way, the surface will be prevented from interaction with the direct laser radiation and the laser-induced plasma, but the effect of peening will be maximized. 

If liquid is used as a confining medium, formation of bubbles and surface waves should be avoided since both may scatter, deflect or defocus the laser beam which leads to irregular ablation of the sacrificial material and therefore non-uniform peening of the sample. However, a universal recipe how to optimize the confining medium and the sacrificial layer to achieve maximal peening efficiency is a subject for future investigations.  

\section{Acknowledgment}

Authors are thankful to Dr. Klaus Neuking, Chair for Materials Science and Engineering, Institute for Materials at Ruhr-Universit\"at Bochum for his support. Lalit M. Kukreja acknowledges financial support received from the Alexander von Humboldt Foundation, Germany under grant no. IND/1015352 for his visiting position at Applied Laser Technologies, Ruhr-Universit\"at Bochum in Germany. Jan Hoppius acknowledges financial support of DFG, Project GU 1075/8 in SPP 1839 {\it Taylored Disorder}.

%
%

\end{document}